\documentclass[aps,prb,twocolumn,amsmath,amssymb,superscriptaddress,floatfix]{revtex4}
\pdfoutput=1
\usepackage[colorlinks=true,linkcolor=blue,citecolor=blue,urlcolor=blue]{hyperref}

\usepackage{graphicx}
\usepackage{graphics}
\usepackage[export]{adjustbox}
\usepackage{bm}
\usepackage{amsfonts}
\usepackage{amssymb}
\usepackage{amsmath}
\usepackage{wasysym}
\usepackage{bbold}
\usepackage{stmaryrd}
\usepackage[usenames]{color}
\usepackage{colordvi}
\usepackage{units}
\usepackage{bbm}
\usepackage{booktabs}
\usepackage{array}
\usepackage{placeins}
\usepackage{epsfig}
\usepackage{lscape}
\usepackage{changes}
\usepackage{braket}
\usepackage{tabularx}
\newcolumntype{L}[1]{>{\raggedright\arraybackslash}p{#1}} % linksbündig mit Breitenangabe
\newcolumntype{C}[1]{>{\centering\arraybackslash}p{#1}} % zentriert mit Breitenangabe
\newcolumntype{R}[1]{>{\raggedleft\arraybackslash}p{#1}} % rechtsbündig mit Breitenangabe

\newcommand{\be}{\begin{equation}}
\newcommand{\ee}{\end{equation}}
\newcommand{\beqn}{\begin{eqnarray}}
\newcommand{\eeqn}{\end{eqnarray}}

\bibstyle{apsrev.bib}

\begin{document}

\title{Statistics of percolating clusters in a model of photosynthetic bacteria}
\author{Jean-Christian Angl\`es d'Auriac}
\email{dauriac@neel.cnrs.fr}
\affiliation{Institut N\'eel-MCBT CNRS, B. P. 166, F-38042 Grenoble, France}
\author{Ferenc Igl{\'o}i}
\email{igloi.ferenc@wigner.hu}
\affiliation{Wigner Research Centre for Physics, Institute for Solid State Physics and Optics, H-1525 Budapest, P.O. Box 49, Hungary}
\affiliation{Institute of Theoretical Physics, Szeged University, H-6720 Szeged, Hungary}
\date{\today}

\begin{abstract}
In photosynthetic organisms, the energy of light during illumination is absorbed by the antenna complexes, which is transmitted by excitons and is either absorbed by the reaction centers (RCs), which have been closed in this way, or emitted by fluorescence. The basic components of the dynamics of light absorption have been integrated into a simple model of exciton migration, which contains two parameters: the exciton hopping probability and the exciton lifetime.  
During continuous radiation with light the fraction of closed RCs, $x$, continuously increases and at a critical threshold, $x_c$, a percolation transition takes place. Performing extensive Monte Carlo simulations we study the properties of the transition in this correlated percolation model. We measure the spanning probability in the vicinity of $x_c$, as well as the fractal properties of the critical percolating cluster, both in the bulk and at the surface. 
\end{abstract}


\maketitle

\section{Introduction}
In living organisms the conversion of (sun)light to chemical energy is performed during photosynthesis. In this process the absorption of photons takes place at the antenna complexes followed by funneling the excitation energy (exciton) to a specially organised dimer ($P$) in the reaction centers (RC)\cite{Mirkovic_et_al_2016}. During the chemical reaction $P \to P^+$, when the RC becomes "closed", the energy of the exciton is transferred to chemical energy\cite{Maroti_and_Govindjee_2016,Maroti_2019a,Maroti_2019b}. Another excition visiting the closed RC can be redirected to an open RC and the exciton is able to visit several RCs during its lifetime. The search for utilization of the exciton by photochemistry (charge separation) competes with loss by fluorescence emission. To describe theoretically this complicated cooperative process several simplified models have been introduced and studied.\cite{Franck_and_Teller_1938,Niederman_2016,Vredenberg_and_Duysens_1963,Joliot_et_al_1973,Paillotin_1976,Bennett_2018,Lavergne_and_Trissl_1995,de_Rivoyre_et_al_2010}

In the first class of models homogeneous distribution of the closed (and open) RCs is assumed together with a small set of reaction rates\cite{Trissl_1996,de_Rivoyre_et_al_2010}. Within this Joliot-Lavergne-Trissl model the time-dependence of the concentrations of $P^+$ are results by solution of set of ordinary differential equations. However, recent experiments\cite{Maroti_2020} on the time-dependence of the fluorescence yield show important deviations from the predictions of this standard theory. In a second class of models the exciton migration is treated in a more accurate way\cite{Paillotin_1976,Den_Hollander_et_al_1983,Fassioli_et_al_2009,Sebban_and_Barbet_1985,Amarnath_2016,Chmeliov_2016}. The disadvantage of these models that the treatment of the equations needs numerical methods and the final results are often not directly related to measurable observables.

Recently, a simple model is proposed by one of us to describe the migration of the excitons and the closure of the RCs\cite{Maroti_2020}. (Hereafter we refer to it as Exciton Migration (EM) model.) This model goes beyond the homogeneous kinetic model and accounts for the local topology of the RC-lattice as well as the bunching effect of the closed RCs during light excitation (induction). Having different reasonable approximations (standard mean-field, lattice mean-field, cluster mean-field methods) the EM model has been solved analytically and it was possible to compare the theoretical predictions with the experiments. 

Despite its possible experimental relevance, the EM model also has several theoretical challenges. One of those is the question about the structure of the closed RCs. During induction the fraction of closed RCs, denoted by $x$, is continuously increasing with time $t$. The bunching of closed RCs is manifested in the non-vanishing form of the near-site connected correlations, which has been studied in Ref.\cite{Maroti_2020}. With increasing time, i.e. with increasing value of $x$ the typical size of clusters of closed RCs increases and at a critical concentration, $x_c$, a percolation transition takes place. In this paper we ask the question how do the parameters of the model (the hopping probability, $p$, and the life-time of the exciton, $n$) influence the properties of the percolation transition. We perform extensive Monte Carlo (MC) simulations, calculate the spanning probability for different boundary conditions and spanning rules and in this way we determine the value of the percolation threshold, $x_c$, and the correlation length critical exponent, $\nu$. At the critical point the fractal properties of the giant cluster is studied, both in the bulk and at the surface.

The rest of the paper is organised in the following way. The model and its known properties are introduced in Sec.\ref{sec:model}. Results of MC simulations are presented in Sec.\ref{sec:results} and discussed in Sec.\ref{sec:discussion}.

\section{Exciton migration model}
\label{sec:model}
To define the EM model for photosynthetic bacteria we use the concept of photosynthetic units (PSU)\cite{Franck_and_Teller_1938}. In a simplified picture each PSU has a reaction center and a light-harvesting antenna. In the EM model the PSUs are assumed to occupy the sites of a lattice: $i=1,2,\dots, N$. For simplicity the lattice is expected to be regular, having a coordination number $z$, in the present case we consider the square lattice. To each RC a two-state variable $\sigma_i=\{0,1\}$ is assigned, being $\sigma_i=0$ for the open and $\sigma_i=1$ for the closed RC. The control parameter of the process is the fraction of closed RCs, which is defined as $x=\langle \sigma \rangle=\lim_{N \to \infty}1/N \sum_{i=1}^N \sigma_i$, what we will also call occupancy in the following. The control parameter is time dependent, $x=x(t)$, and monotonously increases with $t$. At the starting point all RCs are open, thus $x(0)=0$, and after sufficiently long time all RCs became closed, $x(t)=1$ for $t>t^*$.

\bigskip
\underline{Relaxation process}
\bigskip

In the relaxation process we start from a fully-closed state, and after a sufficient period of time, a fraction of the RCs $(1-x)$ will spontaneously reopen. Using an appropriate weak probing light beam, the dynamics of the system can be studied in this case, too. The structure of the closed RCs is different in induction and in relaxation. In the latter process the closed sites are uncorrelated (like in ordinary (inverse) percolation), but in induction there is bunching of near-staying closed sites. This difference in the structure of the closed RC clusters leads to a hysteresis in the fluorescence spectrum as demonstrated recently in Ref.\cite{Maroti_2020}. In this paper in the following we restrict ourselves to the induction process.

%%%%%%%%%%%%%%%%%%%%%%%%% Fig 1 %%%%%%%%%%
\begin{figure}[h!]
%	\vskip -5cm
\includegraphics[width=1.0\columnwidth]{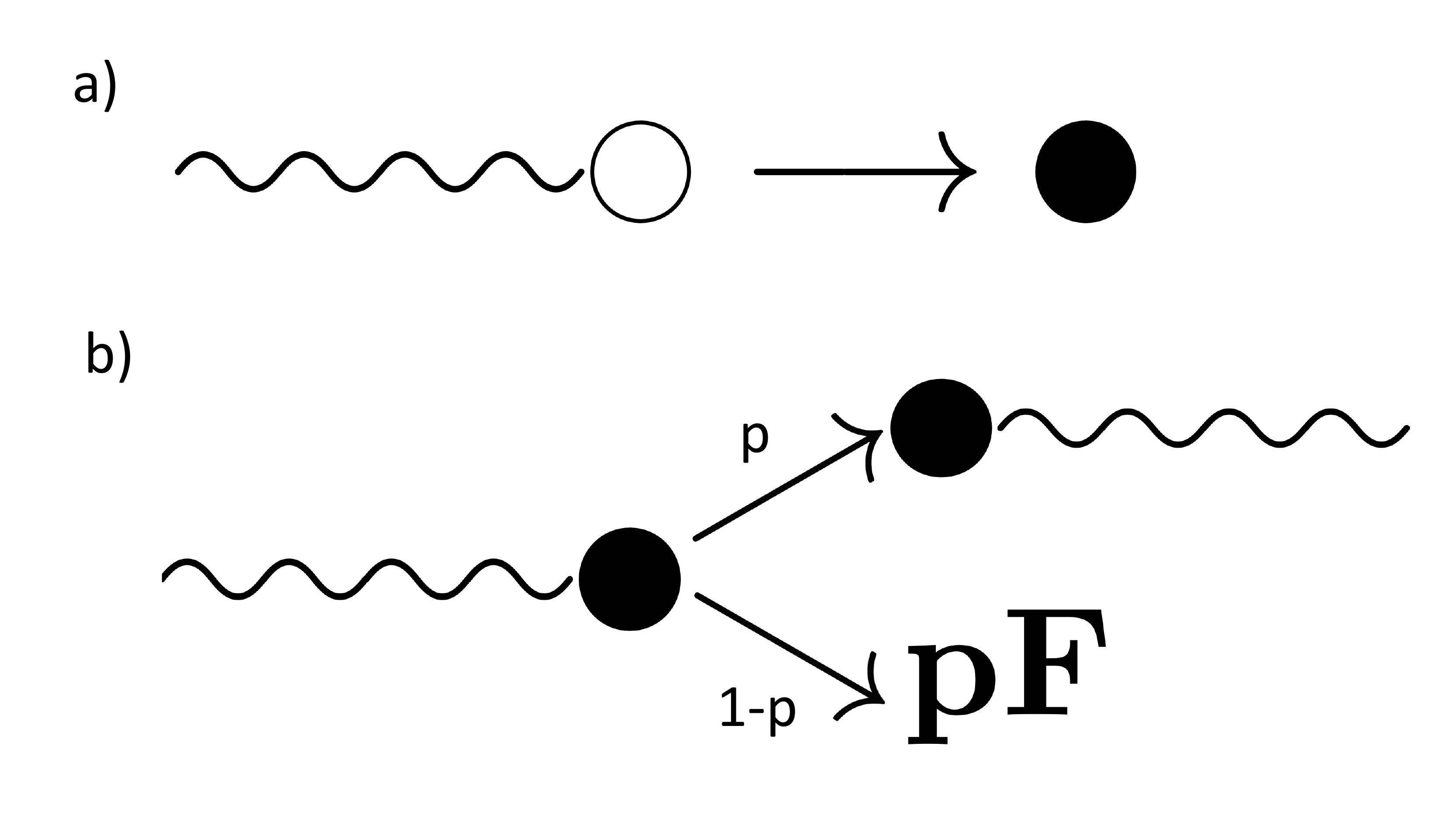}
\caption{Illustration of the processes between an exciton (represented by $\photon$) and an RC. a) If the RC is open (represented by $\fullmoon$) it will become closed (represented by $\newmoon$). b) If the RC is closed, the exciton is redirected with probability $p$, or its energy is dissipated by emission of a fluorescence quantum (represented by \textbf{pF}) with probability $(1-p)$.
\label{fig_1}}	
\end{figure} 
%%%%%%%%%%%%%%%%%%%%%

\bigskip
\underline{Induction process}
\bigskip

In induction, if an incoming exciton hits an open RC it will become closed, see in Fig.\ref{fig_1}a. If, however, the exciton hits a closed RC, two processes can take place. With probability $p$, the exciton is redirected to a neighbouring RC, or with probability $(1-p)$ the energy of the exciton is dissipated by emission of a fluorescence quantum, see in Fig.\ref{fig_1}b. If the exciton is redirected from a closed RC the processes in Fig.\ref{fig_1} are repeated, until the life-time of the exciton. This means that it can visit at most $n$ closed RCs during his wandering. Regarding the exciton wandering we assume that it can only jump to the nearest neighbours, and it can also visit the same closed RC several times.

%%%%%%%%%%%%%%%%%%%%%%%%% Fig 2 %%%%%%%%%%
\begin{figure}[h!]
	\vskip 0cm
\includegraphics[width=1.0\columnwidth]{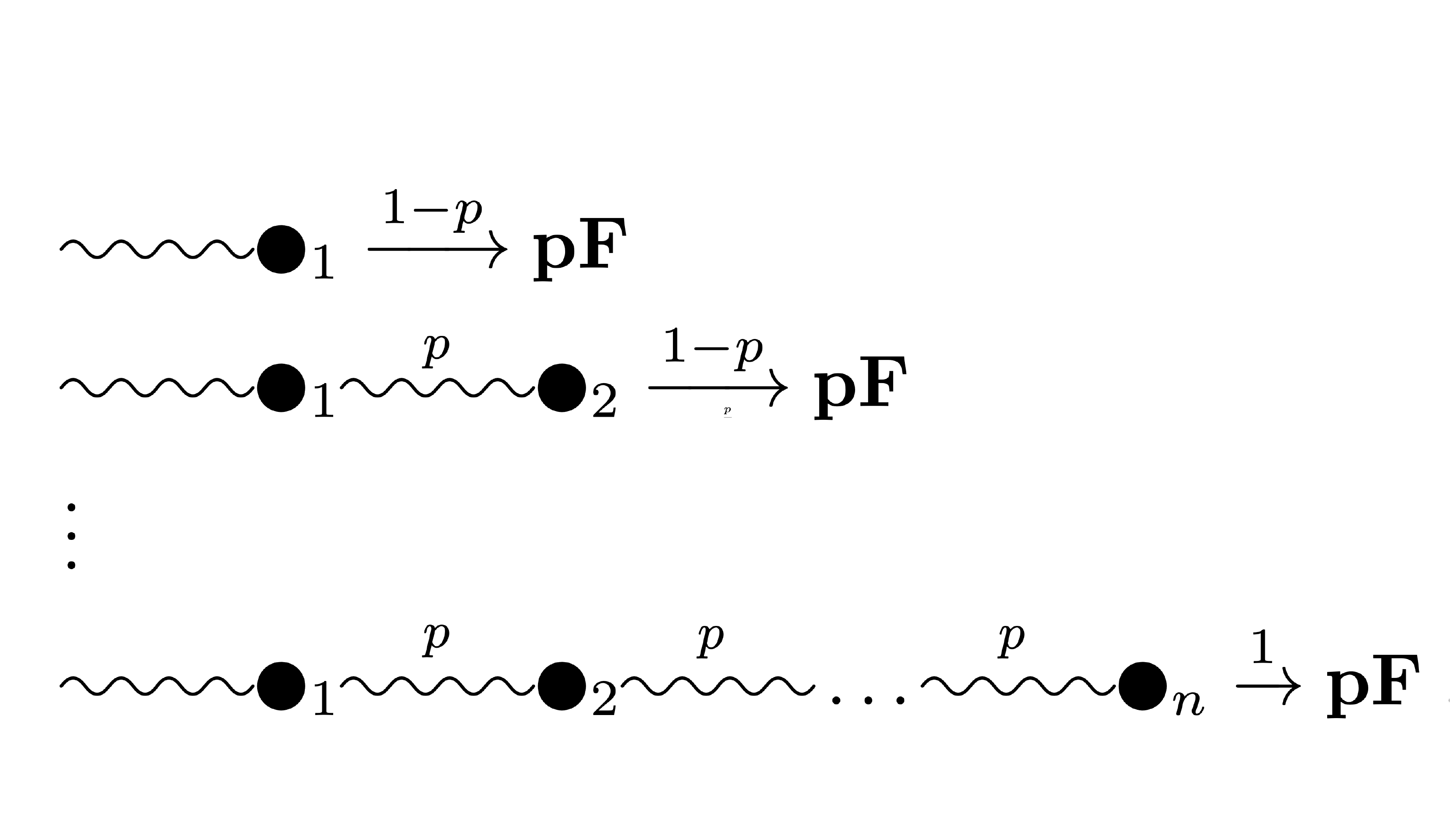}
\caption{Illustration of the processes yielding fluorescence. First line: $1$-step process with a contribution $(1-p)G_1$.
Second line: $2$-step process with a contribution $(1-p)pG_2$. Last line: $n$-step process with a contribution $p^{n-1}G_n$, since the energy of the exciton is radiated with probability one. The $G_k$-s are multisite correlation functions of closed RCs, which are defined in the text. Note, that for $k \ge 3$ step walks the same closed RC can be visited several times.
\label{fig_2}}	
\end{figure} 
%%%%%%%%%%%%%%%%%%%%%

\bigskip
\underline{Fluorescence yield}
\bigskip

The fluorescence yield, $\varphi$, is the sum of the contributions of $k=1,2,\dots,n$ step exciton walks as illustrated in Fig.\ref{fig_2} . This is given by:
\be
\varphi=\sum_{k=1}^{n-1} (1-p)p^{k-1}G_k+p^{n-1}G_n\;,
\label{varphi}
\ee
in terms of the multi-site correlation functions $G_k$. These are the fraction of such $k$-step random walks which visit (nearest neighbour) closed RCs and will be approximated in the following paragraphs, see in Eqs.(\ref{Joliot}),(\ref{lmf}) and (\ref{cmf}). 

Evidently the case $n=1$ or $p=0$ is special, when the exciton is not redirected from a closed RC and the fluorescence yield is given simply as $\varphi=G_1=\langle \sigma \rangle=x$. In this case the structure of clusters follow uncorrelated percolation. However, for $n>1$ (and $p>0$) the multisite correlations are non-zero, $G_k \ne 0$, $k>1$, and there are correlations between the sites of the clusters.

The dynamics of closing the open RCs follows from the fact that all incoming photons that are not emitted through fluorescence will reduce the number of open RCs. Thus the time-dependence of $x$ follows the rule:
\be
\frac{\textrm{d} x}{\textrm{d} t}=1-\varphi\;,
\label{dx/dt}
\ee
where the photochemical rate constant (time-scaling factor) is set $k_I=1$.

\bigskip
\underline{Analytical methods}
\bigskip

In an analytical treatment to integrate the equation in Eq.(\ref{dx/dt}) one needs to use some approximation for the multi-site correlation functions. In Joliot theory \textit{standard mean-field approximation} is used:
\be
G_k \approx x^k,\quad n \to \infty\;,
\label{Joliot}
\ee
so that the exciton can hop to any site and its life-time is unlimited.

The local topology of the lattice, as well as the finite life-time of the exciton is taken into account in the
\textit{lattice mean-field approximation}, in which case the multi-site correlation functions are written in the form:
\be
G_k \approx \sum_{j=2}^k c_j^{(k)}x^j\;,
\label{lmf}
\ee
Here $c_j^{(k)} z^{k-3}$ is the number of $(k \ge 2)$-step random walks, which have visited $2 \le j \le k$ different sites, where the walker arrives to the lattice at the first step. For the square lattice with $z=4$ the first few terms of $c_j^{(k)} z^{k-3}$ are given in Ref.\cite{Maroti_2020}. This approximation is expected to be correct in the relaxation process. 

In induction the bunching of closed RCs is important, which has been taken into account in the \textit{cluster mean-field approximation}. In this case the multi-site correlation functions are approximated in terms of two-site functions, $x_2$, and the density, $x$ as:
\be
G_k \approx \sum_{j=2}^k c_j^{(k)}\frac{x_2^{j-1}}{x^{j-2}}\;,
\label{cmf}
\ee
and $x_2$ has been calculated analytically in Ref.\cite{Maroti_2020}.

\bigskip
\underline{Monte Carlo simulations}
\bigskip

In this paper the time-evolution of $x$ as well as the structure of the closed RC clusters are studied through MC simulations. The RCs are considered to sit on sites of square lattices of size $L \times L$ with $L=2^l$, $l=7,8,\dots,14$ and both free and periodic boundary conditions are used. The external light-source is expected to emit one photon per time-step at discrete times: $t=\tau,2\tau,\dots,m\tau,\dots$ and immediately one exciton starts to move from a randomly selected position of the lattice. Following the rules in Fig.\ref{fig_1} the exciton makes at most $n$ steps during which at most one open RC will be closed. We record the order-parameter, $x(t)$ and the structure of the closed RC clusters, for which the Hoshen–Kopelman algorithm\cite{Hoshen_1976} is used. Particular attention is paid for the properties of the largest cluster. For each size the results are averaged over $10^4$ independent realisations. Since the algorithm
involves the neighbours of a site, the boundary condition
has to be chosen before starting the construction of the sample.
This is in contrast with traditional percolation process (which is the case for $n=1$ or $p=0$) where the boundary condition is necessary only to analyze the samples.

\section{Results about the percolation transition in the EM model model}
\label{sec:results}

\begin{figure}[h!]
	\hspace{0.in}
\includegraphics[width=1.\columnwidth,angle=0]{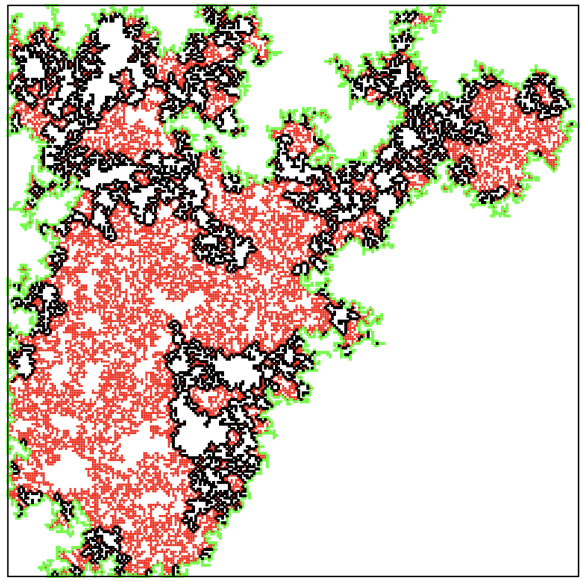}
\caption{Illustration of the fractal structure of the giant cluster for $n=3$ and $p=0.5$ at an occupancy $x=0.589$, slightly above the percolation threshold, see in Table \ref{table1} for $L=256$. Sites in the bulk are shown by red, sites at the perimeter are indicated by green, and the hull is composed from the black and green sites.
\label{fig_3}}	
\end{figure}

As the fraction of closed RCs, $x$, increases connected clusters are formed. These clusters are characterised by a typical mass (number of closed RCs) $m(x)$ and a typical linear size, $\xi(x)$, both monotonously increase with $x$. At a critical value, $x=x_c$, a giant cluster is formed and its size becomes divergent as: $\xi(x) \sim (x_c-x)^{-\nu}$, with $\nu$ being the correlation length critical exponent\cite{Stauffer_1994}. At the critical point the giant cluster is a fractal, its mass is related to its linear size as $m(x_c) \sim \xi^{d_f}$, and $d_f$ is the fractal dimension. The fractal structure of the giant cluster is illustrated in Fig.\ref{fig_3}.
%%%%%%%%%%%%%%%%%%%%%%%%% Fig 3 %%%%%%%%%%
%%%%%%%%%%%%%%%%%%%%%

Above the percolation threshold, $x>x_c$, the giant cluster contains a finite fraction of sites, $P(x)>0$, which behaves in the vicinity of the transition point as $P(x)\sim (x-x_c)^{\beta}$, with a critical exponent $\beta$, the value of which follows from scaling theory\cite{Stauffer_1994} as $\beta=(d-d_f)\nu$. In $2d$ traditional percolation the critical exponents are\cite{Stauffer_1994}:
\be
\nu=4/3,\quad d_f=91/48,\quad \beta=5/36\;.
\label{perc_exp}
\ee
We note that several other processes belong to this universality class, such as invasion percolation\cite{Wilkinson_1983}. Here we aim to study the percolation transition of the EM model and to compare its properties with those of traditional percolation.

\subsection{Spanning probability and critical threshold}
\label{sec:spanning}

We start to estimate the value of the percolation threshold through the calculation of the spanning probability\cite{Roussenq_1976,Reynolds_1978,Reynolds_1980,Eschbach_1981,Cardy_1992,Ziff_1992,Langlands_1994,Aharony_1994,Hovi_1996}, $R(x,L)$, which is the probability that in a given realisation with an occupancy $x$ there is a connected cluster, which span (or percolate) a finite system of linear size, $L$. In the thermodynamic limit the spanning probability behaves as $\lim_{L \to \infty}R(x<x_c,L)=0$ and $\lim_{L \to \infty}R(x>x_c,L)=1$, while at the transition point it has a universal value. For finite systems in the vicinity of the transition point the appropriate scaling combination is\cite{Stauffer_1994}:
\be
\tilde{x}\sim(x-x_c)L^{1/\nu}\;,
\label{scaling_comb}
\ee
and the scaling functions depend on the shape of the system, boundary conditions and spanning rules, but not on the microscopic details of the lattice.

Here we use open boundary conditions in which case three spanning rules have been defined\cite{Reynolds_1978,Reynolds_1980}, which are denoted by ${\cal R}_0$, ${\cal R}_1$ and ${\cal R}_2$. For ${\cal R}_0$ the cluster spans the box either horizontally or vertically, for ${\cal R}_1$ spans the lattice in one given direction (say, horizontally), and for ${\cal R}_2$ spans the box in both directions. For the traditional percolation on the square lattice in addition there are duality relations\cite{Reynolds_1978,Reynolds_1980,Hovi_1996}:
\beqn
R_1(\tilde{x})+R'_1(\tilde{x}')&=&1\;,\nonumber \\
R_0(\tilde{x})+R'_2(\tilde{x}')&=&1\;,
\label{duality}
\eeqn
where the primed quantities are for the dual lattice, for which $\tilde{x}'=-\tilde{x}$.

\begin{figure}[h!]
	\vskip 0cm
\includegraphics[width=1.\columnwidth,angle=0]{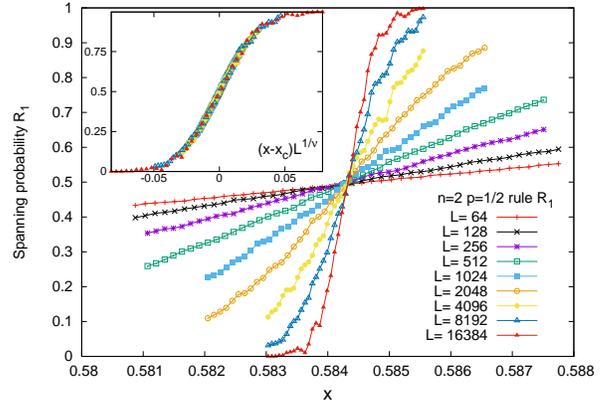}
\caption{Spanning probability of the EM model with $p=0.5$ and $n=2$ on the square lattice with open boundary conditions having the rule ${\cal R}_1$ for different linear sizes $L=64,128,\dots,16384$, up to down. The crossing point of the curves defines the critical point, $x_c=.5843$, where $R_1(x_c,\infty)\approx 0.5$. In the inset the scaling collapse of the points averaged through duality (see text) is plotted in terms of the variable in Eq.(\ref{scaling_comb}). \label{fig_4}}	
\end{figure}

In our numerical work we have calculated the spanning probability of the EM model for different parameters $p>0$ and $n>1$ having the rules ${\cal R}_1$ and ${\cal R}_2$. As an illustration we show in Fig.\ref{fig_4} $R_1(x,L)$ for $p=0.5$ and $n=2$. The curves for different sizes cross each other at about the same point, which defines an estimate for the critical threshold, $x_c=.5843$ and a value for the critical spanning probability: $R_1(x_c,\infty)\approx 0.5$. The measured critical threshold is somewhat lower, than for traditional site percolation, having $x_c=0.5927460$\cite{site_perc,site_perc1}, but $R_1(x_c,\infty)$ agrees with the universal value of traditional percolation: $R_1(x_c,\infty)=1/2$, which follows from duality and has been calculated through conformal invariance\cite{Cardy_1992}. Using the estimate for $x_c$ the points calculated at different sizes can be put to a scaling curve in terms of the variable in Eq.(\ref{scaling_comb}), in which the correlation-length critical exponent for traditional percolation in Eq.(\ref{perc_exp}) is used (not shown). We have also checked the possible validity of the duality relation in the first equation of Eq.(\ref{duality}) and in the inset of Fig.\ref{fig_4} we have plotted the quantity: $(R_1(x-x_c,L)+[1-R_1(x_c-x,L)])/2$ as a function of the scaling variable $\tilde{x}$ in Eq.(\ref{scaling_comb}). As seen in this inset the duality relation is seemingly valid for the EM model and assuming its validity the estimate for the critical threshold could be made more accurate: $x_c=.58432$.

%%%%%%%%%%%%%%%%%%%%%%%%% Fig 4 %%%%%%%%%%
%%%%%%%%%%%%%%%%%%%%%

\begin{figure}[h!]
	\vskip 0cm
\includegraphics[width=1.\columnwidth,angle=0]{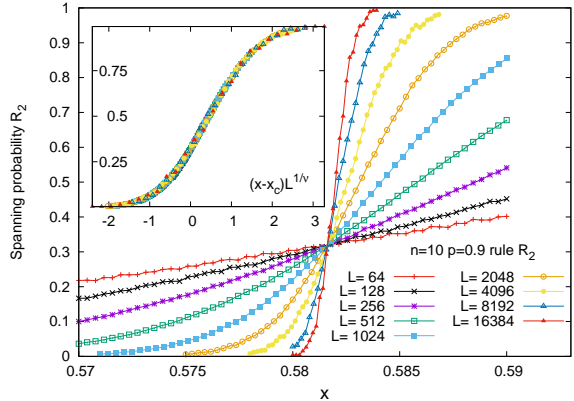}
\caption{Spanning probability of the EM model with $p=0.9$ and $n=10$ on the square lattice with open boundary conditions having the rule ${\cal R}_2$ for different linear sizes $L=128,256,\dots,16384$, up to down. The crossing point of the curves defines the critical point, $x_c=.5817$, where $R_2(x_c,\infty)\approx 0.32$. In the inset the scaling collapse of the points is shown in terms of the variable in Eq.(\ref{scaling_comb}).
\label{fig_5}}	
\end{figure}

To illustrate the use of the spanning rule ${\cal R}_2$ we show in Fig.\ref{fig_5} the spanning probabilities for $p=0.9$ and $n=10$ and for different finite sizes. In this case the crossing point of the curves leads to the estimate: $x_c=.5817$ and the measured value of the critical spanning probability: $R_2(x_c,\infty)\approx 0.32$ agrees well with the conformal result: $0.3223$\cite{Langlands_1994}. Using the estimate for $x_c$ the points calculated at different sizes can be put to a scaling curve in terms of the variable in Eq.(\ref{scaling_comb}), in which the correlation-length critical exponent for traditional percolation in Eq.(\ref{perc_exp}) is used, see in the inset of Fig.\ref{fig_5}.

%%%%%%%%%%%%%%%%%%%%%%%%% Fig 5 %%%%%%%%%%
%%%%%%%%%%%%%%%%%%%%%

Repeating the calculation for another set of parameters of the EM model similar results are obtained. The critical threshold values are found to depend (weakly) on the parameters, but the critical values of the spanning probability is found universal for a given class of spanning rule. The $x_c$ values are collected in table \ref{table1}, together with the fraction of points in the critical cluster which have been closed directly. At a fixed value of $n>1$ the critical threshold monotonously decreases with increasing $p$ and the same is true for the fraction of directly closed points, too. On the other hand at a fixed value of $0<p\le 1$ the $n$ dependence of $x_c$ is non-monotonous,
it has the minimum at $n=2$.  

%%%%%%%%%%%%%%%%%%%%%%% table 1 %%%%%%%%%%%%%%%
\begin{center}
\begin{table}
\begin{centering}
\begin{tabular}{|c|c|c|c|}
\hline 
 & $n=2$ & $n=3$ & $n=10$\tabularnewline
\hline 
\hline 
%$p=0$ & 0.592746 1 & 0.592746 1 & 0.592746 1\tabularnewline
%\hline 
$p=0.1$ & 0.5908 (0.972) & 0.5908 (0.971) & 0.5908 (0.971)\tabularnewline
\hline 
$p=0.5$ & 0.5843 (0.881) & 0.5845 (0.867) & 0.5845 (0.861)\tabularnewline
\hline 
$p=0.9$ & 0.5794 (0.813)  &   0.5796 (0.779)  &  0.5817 (0.746) \tabularnewline
\hline 
$p=1$ & 0.5785 (0.798) & 0.5800 (0.759) & 0.5836 (0.712) \tabularnewline
\hline 
\end{tabular}
\caption{Critical threshold values for different parameters. For traditional site-percolation ($p=0.$ or $n=1$) we have $x_c=0.5927460$\cite{site_perc,site_perc1}. In parenthesis the proportion of closed sites which have been closed directly are indicated.
\label{table1}}
\par\end{centering}
\end{table}
\par\end{center}
%%%%%%%%%%%%%%%%%%%%%%%%%%%%%%%%%%%%%%%%%%%%%%%

\subsection{Fractal dimensions of the critical cluster}
\label{sec:fractal}

The scaling properties of the spanning probability in the previous subsection indicate, that the EM model probably belongs to the universality class of traditional percolation. Both the values of the critical spanning probability and the correlation length critical exponent support this assumption. Here we check further this point and calculate the fractal dimension of the critical cluster, both in the bulk and at the surface.

%%%%%%%%%%%%%%%%%%%%%%%%% Fig 6 %%%%%%%%%%
\begin{figure}[h!]
	\vskip 0cm
\includegraphics[width=1.05\columnwidth,angle=0]{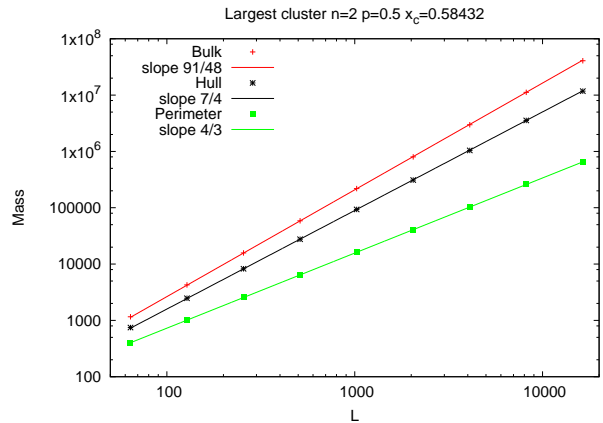}
\caption{Fractal properties of the critical clusters in the EM model with $p=0.5$ and $n=2$ on the square lattice. Average mass, $m(L)$, hull, $h(L)$ and perimeter, $s(L)$, as a function of linear size, $L$, in log-log scale. The slope of the straight lines through the points have the fractal dimensions of traditional percolation in Eqs.(\ref{perc_exp}) and (\ref{fractal_dim_surf}). 
\label{fig_6}}	
\end{figure} 
%%%%%%%%%%%%%%%%%%%%%

In order to determine the bulk fractal dimension, $d_f$, we have calculated the average mass of the critical cluster, $m(L)$ for finite systems of linear size, $L$, which is expected to scale asymptotically as $m(L) \sim L^{d_f}$. In Fig.\ref{fig_6} we plotted $m(L)$ as a function of $L$ in log-log scale for the EM model with $p=0.5$ and $n=2$. As seen in the figure the points are almost perfectly on a straight line the slope of which is compatible with the fractal dimension of traditional percolation in Eq.(\ref{perc_exp}). 

Regarding the surface properties of the critical cluster two quantities are usually studied, the hull and the perimeter\cite{Stauffer_1994}, see Fig.\ref{fig_3} for an illustration. The hull contains sites on the cluster that neighbour open sites, which are connected to the outside. On the contrary accessible external perimeter sites of the cluster are those points which can be hit by an extended diffusing particle from outside with non-zero probability. The average number of sites on the hull and at the perimeter are denoted by $h(L)$ and $s(L)$, respectively. We have calculated both quantities for the EM model with $p=0.5$ and $n=2$ and these are plotted vs. $L$ in log-log scale in Fig.\ref{fig_6}. As seen in this figure the points of both quantities are on straight lines, so that we have asymptotic power-law relations:
\be
h(L) \sim L^{d_h},\quad s(L) \sim L^{d_s}\;.
\ee
In addition the measured slopes of the straight lines are in agreement with the known values of the fractal dimensions for traditional percolation\cite{saleur_1987}:
\be
d_h=7/4,\quad d_s=4/3\;.
\label{fractal_dim_surf}
\ee
\section{Discussion}
\label{sec:discussion}

In this article we examined the properties of percolation clusters formed by closed photosynthetic units of photosynthetic bacteria during light emission. The process is described in the context of an exciton migration model in which nearby sites of the clusters are closed in a correlated manner. This model contains two parameters, the exciton hopping probability, $p$, and the exciton lifetime, $n$. After the sample has been illuminated the fraction of closed sites, $x(t)$ increases in time and the typical (linear) size and mass of the clusters increase, too. At a critical occupancy, $x_c$, a giant cluster is formed and a percolation transition takes place.

In this paper the EM model is considered on the square lattice and the properties of the percolation transition is studies through MC simulations for different values of the parameters. The critical percolation threshold is calculated by studying the spanning probability for open (and periodic) boundary conditions and for different spanning rules, ${\cal R}_1$ and ${\cal R}_2$. The $x_c$ values are found to vary weakly on the parameters and the same is true for the proportion of directly closed sites, see in Table \ref{table1}. On the other hand the critical spanning probability is found to be universal: for a given boundary condition and a spanning rule its value is independent of the parameters of the model and corresponds to the exactly known values for (uncorrelated) traditional percolation.

From the finite-size scaling behaviour of the spanning probability in the vicinity of the critical point the correlation length critical exponent, $\nu$, is extracted and its value is found also compatible with that of traditional percolation.
We have also studied the fractal properties of the critical cluster, in the bulk at the perimeter and in the hull. At each cases the corresponding fractal dimensions are found universal, their values are compatible with the known respective values for traditional percolation.

To summarise the effect of correlations caused by the exciton migration is found to be irrelevant in the critical properties of the percolation transition of the EM model. Consequently these correlations are sufficiently short-ranged and they do not influence the asymptotic large scale correlations in the system. This is found true both for bulk and surface correlations. Similar conclusions are expected to hold for other type of lattices as well as in higher dimensions.

\begin{acknowledgments}
This work was supported by the National Research Fund under Grants No. K128989, No. K115959 and No. KKP-126749. F.I. thanks previous cooperation in the project to P. Mar\'oti, I. A. Kov\'acs, M. Kis and J. L. Smart and valuable discussions with L. Gr\'an\'asy.
\end{acknowledgments}

%\vskip 3cm
%\section*{References:}

\end{document}